\documentclass{aa}
\usepackage{graphicx}
\usepackage{rotating}
\usepackage[round]{natbib}
\bibpunct{(}{)}{;}{a}{}{,}
\begin{document}
%
\title{Detection of Wolf-Rayet stars in host galaxies of Gamma-Ray Bursts (GRBs):
are GRBs produced by runaway massive stars ejected from high stellar density regions ?
\thanks{Based on VLT/FORS2 observations collected at the European Southern 
Observatory, Paranal, Chile, programme No. 073.B-0482(A).} }

\author{ F. Hammer\inst{1}, H. Flores\inst{1}, D. Schaerer\inst{2,3}, 
M. Dessauges-Zavadsky\inst{2}, E. Le Floc'h\inst{4,5}, and M. Puech\inst{1}}

\offprints{francois.hammer@obspm.fr}
\authorrunning{F. Hammer et al.}
\titlerunning{GRB}

   \institute{Laboratoire Galaxies Etoiles Physique et
        Instrumentation, Observatoire de Paris-Meudon, 5 place Jules Janssen,
        92195 Meudon, France
        \and
        Observatoire de Gen\`eve, 51 Ch. des Maillettes, 1290 Sauverny, 
	Switzerland
        \and
        Laboratoire d'Astrophysique Toulouse-Tarbes, UMR 5572, 14, Av. E. Belin,
        31400 Toulouse, France
        \and
        Steward Observatory, University of Arizona, 933 North Cherry Avenue, 
	Tucson, AZ 85721, United States
        \and
        also associated to Observatoire de Paris, GEPI, 92195 Meudon, France
        }
\date{---}

\abstract{
We have obtained 
deep spectroscopic observations of several nearby gamma-ray burst
(GRB) host galaxies revealing for the first time the presence of Wolf-Rayet (WR) stars and
numerous O stars located in rich and compact clusters or star forming regions.
Surprisingly, high spatial resolution imaging shows that the GRBs and the 
associated supernovae did not occur in these regions, but several hundreds of parsec away.
Considering various scenarios for GRB progenitors, we do not find any simple 
explanation of why they should be preferentially born in regions with low 
stellar densities. All the examined GRBs and associated SNe have occurred 400 
to 800 pc from very high density stellar environments including large numbers of 
WR stars.  Such distances can be travelled through at velocities of 100 km~s$^{-1}$ or larger, assuming the travel time to be the typical life time of WR stars. It leads us to suggest that GRB progenitors may be 
runaway massive stars ejected from compact massive star clusters. 
The ejection from such super star clusters 
may lead to a spin-up of these stars,
producing the loss of the hydrogen and/or
helium envelopes leading to the origin of the type Ibc
supernovae associated with GRBs. If this scenario applies tocd text/Sc	
all GRBs, it provides a natural explanation of the very small fraction
of massive stars that emit a GRB at the end of their life.  An alternative to this scenario could be a binary origin for GRBs, but this still requires an explanation of why it would preferentially occur in low stellar density regions.
\keywords{cosmology: observations, galaxies: individual (GRB980425, GRB020903, 
GRB031203), galaxies: stellar content, galaxies: abundances, Stars: Wolf-rayet}
}

\maketitle
%

\section{Introduction} 

Gamma-Ray Bursts (GRBs) are believed to trace the death of massive,
short lived stars, providing the most energetic events in the
Universe. This is further supported by the discovery that several
long-duration GRBs are, indeed, associated with the collapse of
massive stars to a black hole, referred to as the collapsar model 
\citep{galama98,stanek03,hjorth03}. In this
model, a rapidly rotating star undergoing core-collapse produces a
jetted GRB along the rotation axis, and blows up the entire star in an
energetic supernova explosion \citep{macfadyen99,klose04}. 
The association between GRBs and supernovae (SNe)
indicates in many cases that the parent SN population of GRBs is
formed by peculiar type Ibc SNe which have been proposed to be
SN1988bw-like SNe by \citet{foley03}. They are characterised by
high luminosity peaks ($M_B$ from $-19.5$ to $-17$) and a high expansion
velocity of the ejecta (30\,000 km~s$^{-1}$). Nevertheless, other types of
SNe (more regular SNIbc or even SNIIn) cannot be excluded. The
identity of the progenitors of GRBs has thus still to be
addressed. However, while the peculiar SNIbc may be further confirmed as
the parent SN population of GRBs, the most favoured progenitors for
the collapsar model are the Wolf-Rayet (WR) stars.

 \citet{hirschi05} have investigated WR stars
and have shown that stellar evolution models including detailed
effects of rotation predict the conditions for GRB production
via collapsars, namely the conditions for black hole
formation, the loss of the hydrogen-rich envelope (such as in SNIbc)
and enough angular momentum to form an accretion disk around the black
hole. Furthermore, if only stars of the particular WO subtype --
thought to be intimately related to SNIc -- are considered, the GRB
production rate can fairly well be reproduced.
In their model, GRBs are predicted to occur only over a limited
metallicity interval at subsolar values (typically at $Z_{\rm SMC}$ to
$Z_{\rm LMC}$, i.e.\ $Z \sim 0.2-0.4$~$Z_\odot$).  However, when
magnetic fields are taken into account in these models, it may well
be more difficult to produce GRBs \citep{petrovic05}. To
circumvent this problem \citet{woosley06} and \citet{yoon05} 
suggest rapid rotators at low metallicity (typically $Z \la 0.05$~$Z_\odot$) 
as GRB progenitors, since in these cases a nearly
chemically homogeneous evolution and a low stellar mass loss can produce
the right conditions (high specific angular momentum stars with no
hydrogen envelope) for the collapsar model. Establishing a precise
metallicity limit, below which this scenario may work, is however
difficult due to uncertainties in stellar mass loss rates and initial
stellar rotation rates \citep{yoon05}.
To establish observationally the high mass of long-duration GRB
progenitors and the collapsar model, and to constrain scenarios like
those just mentioned, it is imperative to confirm the presence of WR
stars in the region of the GRB, which has never been achieved until
now, and to better determine progenitor metallicities.

The first complete optical study of $z < 1$ host galaxies has been
done by \citet{lefloch03}. It revealed that GRBs occur in
galaxies with low luminosities and blue colors 
\citep[see also][]{fruchter99,sokolov01}. 
In addition, there is growing evidence that the majority
of GRB host galaxies are Ly$\alpha$ emitters with star formation rates
(SFRs) between 1 to 11 M$_{\odot}$/yr \citep{fynbo03,jakobsson05}. 
All this is indicative of low metallicity environments,
as confirmed by direct abundance measurements (Prochaska et~al. 2004; 
Sollerman et~al. 2005; Hammer et~al., in preparation).
As a result, it leads to some controversy, since GRB hosts were
formerly believed to be associated with galaxies with strong star
formation rates averaging 100 M$_\odot$/yr from radio and sub-mm
observations \citep{berger03}. It is unclear if the radio emission
is simply related to star formation or if the sub-mm detection could
be affected by the lack of spatial resolution: this may be questioned
by the absence of detection of most GRB hosts by Spitzer \citep{lefloch06}.

The analysis of the overall spectral energy distribution
(SED) shows that the GRB hosts have high specific star formation rates
(i.e.\ high SFRs with respect to their luminosity) and younger stellar
populations than an ensemble of all field galaxies \citep[e.g.][]{christensen04}. 
However, as these authors note, the derived ages of
 $50-200$ Myr seem to indicate that GRB hosts are not
significantly younger than starburst galaxies at similar
redshifts. These ``old'' ages measured could be the result of large
aperture effects or composite stellar populations, i.e. the young
population from which the GRB descends might be diluted by older
stars.

Further understanding of the GRB production mechanism requires us to
study their environments in detail. In this paper we present deep
spectroscopic observations of GRB hosts, and focus our analysis on the
four most nearby host galaxies: GRB980425 at $z=0.008$, GRB031203 at
$z=0.1055$, GRB030329 at $z=0.169$ and GRB020903 at $z=0.25$.  They
are bright enough to test for the presence of WR stars, as well as to
robustly establish metal abundances through a direct electron
temperature measurement.
Moreover, they are found to be associated with a Type Ibc SN
(GRB980425/SN1998bw: Patat et~al. 2001; GRB031203/SN2003lw: 
Malesani et~al. 2004; GRB030329/SN2003dh: Hjorth et~al. 2003;
GRB020903: Soderberg et~al. 2005), and when using HST imaging, the SN
location can be identified with some accuracy within the host galaxy.

Our VLT observations are described in Sect.\ 2.
The properties of the environments of GRB980425 and GRB020903 as deduced
from spectroscopy and imaging are discussed in Sect.\ 3.
Based on this data we suggest in Sect.\ 4 a new scenario for GRB progenitor
stars. In Sect.\ 5 we summarise our result and discuss possible implications.
Throughout the paper
we adopt the $\Lambda$CDM cosmological model ($H_0$=70 km~s$^{-1}$
Mpc$^{-1}$, $\Omega _M$=0.3 and $\Omega _\Lambda =0.7$).

\section{Observations and measurements}

Spectroscopic VLT/FORS2 observations were done in visitor mode in July
2004 (programme No. 073.B-0482(A)). Among our eight targets, two
nearby galaxies were observed, the GRB980425 at $z=0.0085$ and the
GRB020903 at $z=0.25$ using two differents FORS2 set-ups (600B and
600RI grisms with a resolution $R\sim 1300$). For the GRB980425 host, 
given the size of the galaxy, the slit was placed as it is shown in Fig.~1. 
The GRB980425 host was observed with a total
exposure time of 1800s and 1500s with the 600B and 600RI grisms,
respectively, and the GRB020903 host 7200s with each grism.

Data reduction and extraction of optical spectra were performed using
a set of IRAF procedures developed by our team, which allowed us to
reconstruct simultaneously the spectra and the sky counts of the
objects. Spectrophotometric calibration of each grism was done using
the same star. Broadband filter images were used to compute aperture
corrections and check the spectrophotometric calibration.

\begin{table}
\caption{Flux measurements of the GRB host galaxies}
\begin{tabular}{lrrrr}
                & \multicolumn{3}{c}{GRB980425} & GRB020903 \\
                & SN reg.  & WR reg. & reg. 4   &           \\\hline
\ion{O}{II}   $\lambda$3727    & 4.020    &  44.00  & 2.680    &  6.89  \\
\ion{Ne}{III} $\lambda$3869   & 0.550    &  11.30  & 0.055    &  2.05  \\
\ion{O}{III}  $\lambda$4363   & $<$0.030 &   1.55  & $<$0.030 &  0.54  \\
\ion{N}{III}  $\lambda$4640   &   --     &   0.30  &   --     &  0.0078 \\
\ion{He}{II}  $\lambda$4686   & $<$0.052 &   0.53  &   --     &  0.026 \\
\ion{Ar}{IV}  $\lambda$4711   &   --	  &   0.34  &   --	&  0.12  \\
\ion{Ar}{IV}  $\lambda$4740   &   --	  &   0.16  &   --	&  0.20  \\
H$\beta$                      & 0.910    &  35.40  & 0.230	&  4.40  \\
\ion{O}{III}  $\lambda$4959   & 0.900    &  49.70  & 0.570	&  7.40  \\
\ion{O}{III}  $\lambda$5007   & 2.040    & 201.20  & 1.100	& 33.50  \\
\ion{N}{II}   $\lambda$5755   &   --     &   0.17  &   --     &   --   \\
\ion{O}{I}   $\lambda$6300     & 0.400    &   1.70  & 0.200    &  0.65  \\
\ion{S}{II}  $\lambda$6312    & 0.040    &   0.90  &   --	&   --   \\
\ion{N}{II}  $\lambda$6548    & 0.380    &   4.17  & 0.340	&  0.25  \\
H$\alpha$                     & 3.860    & 183.90  & 1.890    & 16.80  \\
\ion{N}{II}  $\lambda$6583    & 0.880    &  11.30  & 0.520	&  0.72  \\
\ion{S}{II}  $\lambda$6716    & 0.600    &  10.90  & 0.540	&  2.39  \\
\ion{S}{II}  $\lambda$6731    & 0.840    &   8.60  & 0.590	&  1.71  \\
\ion{Ar}{III}  $\lambda$7136  & 0.190    &   7.90  & 0.096	&   --   \\
\ion{O}{II}  $\lambda$7325    & 0.195    &   3.20  & 0.126    &   --   \\
\ion{Ar}{II} $\lambda$7751  & 0.03     &   2.08  &   --	&   --   \\
\end{tabular}
\begin{description}
\item [Notes:] Measured fluxes before aperture correction in 10$^{-16}\times$ 
erg$^{-1}$~cm$^{-2}$~\AA$^{-1}$. Typical errors (1$\sigma$) are 0.03 for the faint lines
as estimated from the SPLOT routine.
\end{description}
\end{table}

Flux measurements were performed using the SPLOT package of IRAF.
Measurements were performed by two of us (F.H. and H.F.) and
compared. In the case of GRB980245 we also compared our results
with those performed on spectra for which the background light of the
ESO 184-G82 galaxy had been removed. All results are found to be very
similar, which is supported by the remarkable consistency of the derived
temperatures using different chemical species. The electron density
was derived from the S~{\sc ii} line ratio, the electron temperature
from O~{\sc ii} $I(3726+3729)/I(7320+7330)$, O~{\sc iii} $I(4959+5007)/I(4363)$, 
N~{\sc ii} $I(6548+6583)/I(5755)$, and S~{\sc ii} $I(6716+6731)/I(4068+4076)$, when
available. These estimates were made using the NEBULAR package of
IRAF, after correcting the lines for extinction. The package is based
on a 5-level atom program and is described by \citet{shaw94}. 
Extinction correction was estimated using the
H$\alpha$/H$\beta$ ratio, and verified using higher order Balmer
lines. All have been found in good agreement, but one, for which we
have corrected fluxes from the underlying absorption using the method
of \citet{izotov99} based on H$\alpha$/H$\beta$ and
H$\gamma$/H$\beta$ ratios.

\begin{table*}
\caption{Physical properties of the GRB host galaxies}
\begin{tabular}{lrrrr}
                 & \multicolumn{3}{c}{GRB980425}& GRB020903 \\
                 & region SN& region WR & region 4 &  	 \\\hline
N$_{e}$ [S~{\sc ii}]$^{a}$  &    27    &	158  &  498  &  27    \\
T$_{e}$ [O~{\sc ii}]$^{b}$  &    11939 &  11129 &  9160 &  --    \\
T$_{e}$ [O~{\sc iii}]$^{b}$ &    $<$13966 &  11900 & $<$18274 &  14852 \\
T$_{e}$ [N~{\sc ii}]$^{b}$  &      --  &  11249 &   --  &  --    \\
12+log(O/H)]$^{c}$      &    8.25  &  8.39  &  9.00 & 7.97   \\
N/O              &    0.21  &  0.043 &  --   & 0.013  \\
A$_V$$^{d}$           &    1.02  &  1.51  &  0.98 &  0.76  \\
\end{tabular}       
\begin{description}
\item [Notes:] (a)~Electronic density estimated from the S~{\sc ii} flux
ratio $I(6716)/I(6731)$. (b)~Temperature in K estimated from the
following flux ratios: O~{\sc ii} $I(3726+3729)/I(7320+7330)$, O~{\sc iii}
$I(4959+5007)/I(4363)$, N~{\sc ii} $I(6548+6583)/I(5755)$, and S~{\sc ii}
$I(6716+6731)/I(4068+4076)$. (c)~Metallicity, 12+log(O/H), estimated using the
effective temperature method. (d)~Extinction coefficient, A$_V$ (in magnitudes), has been computed using
the standard Balmer ratio of H$\beta$ and H$\alpha$.
\end{description}
\end{table*}

\section{Properties of the environments of GRB980425 and GRB020903}


We now discuss our FORS2 spectra of the SN1998bw remnant and several nearby
regions of the GRB980425 host as well as HST imaging of this galaxy, and
FORS2 spectroscopy and HST imaging of the more distant GRB020903.  A
brief description of the hosts of GRB030329 and GRB031203 (based on
archival data) is given for comparison in Sect.~4.

\subsection{GRB980425 host galaxy}

Figure~1 displays the slit position with labels for the regions around
SN1998bw. HST/STIS
images (ESO/HST archive programmes 8640, 8243, and 8648) of two
regions are shown, revealing 7 sub-areas in the immediate vicinity of
SN1998bw \citep{fynbo00}, and the ``WR region'' includes a very
luminous main component, hardly resolved (0.03 arcsec FWHM), corresponding
to 5~pc FWHM.

\begin{figure*}
\centering 
\caption{SEE ATTACHED JPG FIGURE - VLT image of the ESO-184-G82 galaxy showing the slit location and
	the regions surrounding the SN1998bw remnant. On the right,
	enlargement of the two area (WR region on top, SN region on
	bottom) based on HST/STIS imagery. The location of the SN is indicated by a circle.}
\end{figure*}

\subsubsection{A bright and compact HII region with WR stars} 

This region lies approximately 0.8 kpc NW from the SN1998bw, and
its spectrum presents numerous emission lines, with a broad range
of chemical species (see Fig.~2 and Table~1). Its density, temperature
and oxygen abundance (see Table~2) are characteristic of a vigorous
star forming region with $Z \sim 0.5$~$Z_{\odot}$ (assuming 12+log(O/H) =
8.69 for the solar abundance, see Allende Prieto et al., 2001).  H$\alpha$ and H$\beta$ show very large
equivalent widths, 1400 and 200 \AA, respectively. After correcting
for the aperture effect (factor 4.6) and for extinction, we find a
luminosity of $1.24\times 10^{40}$ erg~s$^{-1}$ in the H$\beta$ line,
which corresponds to $\sim$ 2300 equivalent O7V stars 
\citep[cf.][]{schaerer98}.

\begin{figure*}
\centering
\includegraphics[width=15cm]{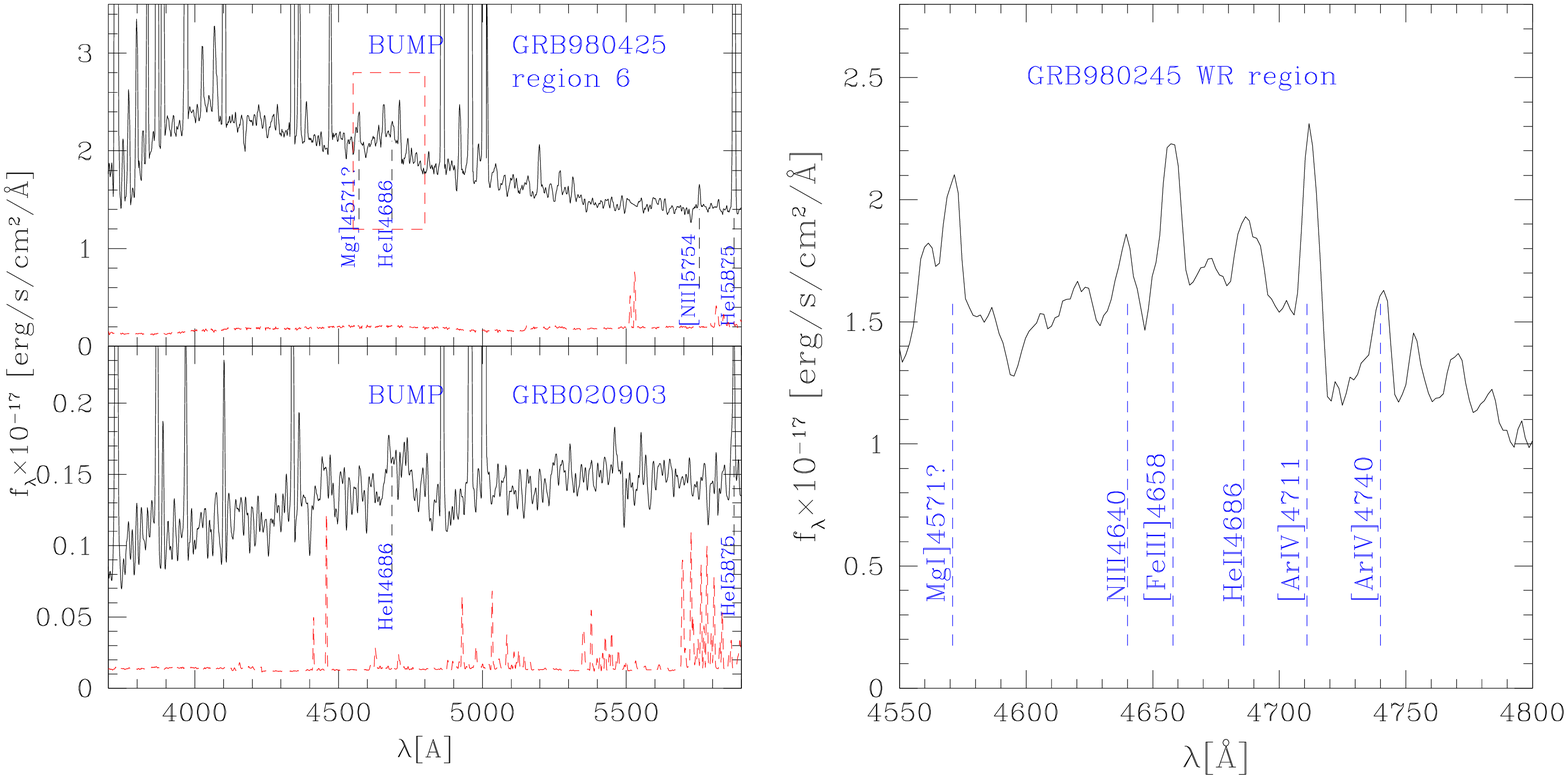}
\caption{Spectra of the WR region of the GRB980425 host (top, left)
	and of the GRB020903 host (bottom, left). On the right is
	shown an enlargement around He{\sc ii}4686 of the spectrum of the WR
	region of the GRB980425 host.}
\end{figure*}

The spectrum shows a prominent blue WR  bump around the He~{\sc ii}4686 line,
with many faint emission lines (Ar{\sc iv}4711,4740, N{\sc iii}4640, and
Fe{\sc iii}4658, see Fig.~2). The absence of carbon lines (C{\sc iii}5696 and C{\sc iv}5808)
suggests that most of the WR stars are of the WN type. Using the luminosity
of the blue bump from 4650 to 4686 \AA\ \citep[see][]{schaerer98},
we estimate the number of WN stars to be $86\pm 20$ (1$\sigma$ error
estimated from the continuum) in that region (assuming WN7 star
luminosities). It agrees with the measurement of the faint He{\sc ii}4686
line (see Table~1), which corresponds to the emission of 46 to 92 WN
stars.
The large H$\alpha$ and H$\beta$ equivalent widths, the presence of WR stars, and fits to
the spectral energy distribution are all indicative of a very young
($\sim 1-6$ Myr, depending on the age indicator used) massive star
forming region.  These properties, as well as the size of the region
\citep{oconnell94}, show that we are dealing
with a (possibly multiple?) massive, young, super star cluster.

We also notice the possible presence of MgI]4571 in emission, 
a feature commonly seen in the nebular phase of type Ibc SNe 
\citep[see][]{patat01,foley03} and broad wings in the bottom of the
H$\alpha$ emission line revealing gas motion at velocities up to 3000
km~s$^{-1}$. If real, these features, combined with the large number
of WR stars, might indicate the presence of recent supernovae in this
region.

While the location of SN1998bw is beyond doubt \citep[see][]{galama99}, 
its association with GRB980425 has been much debated. The
major argument in favour of the association of the two events is their
remarkable temporal coincidence.  The fact that other SN of the same
type as SN1998bw have been found a few weeks after GRB events 
\citep[see e.g.][]{foley03} is also a good argument to support the
association. Nevertheless, the peculiar properties of the WR region
near the SN has prompted us to investigate whether some supernovae might
have occurred there few years before our observations. We have used
the HST/STIS imagery \citep[see][]{fynbo00} taken at
three different epochs, namely June and November 2000 and August
2001. Special care has been taken to derive a common astrometric
solution for each image and we derive a 0.03 STIS pixel (or 0.75~mas)
accuracy using three stars around the WR and SN regions. While the
variability of SN1998bw is well detected, we found no indication of
variability in the WR region. We notice, however, that the
detection of a faint signal in a very bright area is prevented by the
Poisson noise of the total signal, and so we
would not be able to see a faint signal such as the SN1998bw remnant two
years after its maximum, if it was embedded in the bright WR region.
However, we believe that a strong variability similar to that of
SN1998bw at its peak would have been detected in the WR region, since
this field has been thoroughly surveyed since May 1998. Recall that
the SN1998bw luminosity near its maximum surpasses the luminosity of
the WR region \citep[see Fig.~2 of][]{galama99}.
Hence, we conclude that the association of the GRB980425 with SN1998bw
is very probable, and that the WR region is a strong star-forming region including a very bright
(possibly multiple) super star cluster separated by 800~pc from the SN.

 The properties of this cluster are indeed interesting, and sufficiently strong to
explain reasonably well the detection at 24~$\mu$m reported by
\citet{lefloch06}.  Indeed, assuming standard SFR conversion
factors, the star formation rate derived from the H$\alpha$ line is  
$0.34$~$M_{\odot}$/yr (aperture and extinction corrected), a value in
excellent agreement with what can be inferred from its IR luminosity
\citep[$L_{IR} = 2\times 10^{9}$~$L_{\odot}$, see][]{lefloch06},
i.e. SFR $= 0.35$~$M_{\odot}$/yr.
>From its absolute scales the properties of the WR region are also
 similar to those of the embedded super star cluster
(``supernebula'') in the nearby starburst NGC 5253.
The bolometric luminosity of the dominant embedded cluster in
NGC 5253 is $L_{\rm bol} \sim (1-3) \times 10^9$~$L_\odot$ 
\citep{beck96,vanzi04}, and it contains several thousand O
stars within a small (pc or even sub-pc scale) region \citep[e.g.][]{turner03}. 
However, the extinction in the WR region of the GRB980425
host is much smaller. Further comparisons are beyond the
scope of this paper.


\subsubsection{The region surrounding the SN1998bw remnant} 

This region is almost 10 times fainter than the WR region, and
provides a spectrum with a lower S/N. Nevertheless, it shows enough
emission lines to derive a full diagnosis of its interstellar medium.
It shows a more moderate extinction and with $Z\sim 0.36$~$Z_{\odot}$
a somewhat lower oxygen abundance than the WR region and region 4 (see Table~2).
%
%
Interestingly, and in contrast to the WR region, this region also
has a high nitrogen over oxygen abundance of
N/O $=0.24$, almost twice the solar value. While such a high N/O ratio is approximately 
4 times larger than that found in HII regions
\citep[cf.\ e.g.][]{liang05,izotov06}, such strong N line
intensities can be found in SN remnants \citep{smith93}.
This N excess
could be related to the enrichment from the progenitor star of SN1998bw, as 
expected from rotating stars. For example, similar N/O ratios are predicted
by the yields of $\sim$ 10--20 M$_\odot$ stars in the rotating stellar evolution 
models of \citet{hirschi05} albeit at somewhat larger metallicity than the
one observed for this SN region. However, dilution with pre-existing ISM will
reduce the resulting N/O ratio.


The SN region shows moderate H$\alpha$ and H$\beta$
equivalent widths of 88 and 16 \AA, respectively,
 indicative of ages of $\sim$ 6--8 Myr adopting our evolutionary synthesis models.
The aperture
(factor 3.27) and extinction corrected H$\beta$ flux corresponds to
only $\sim$ 10 O7V equivalent stars.
We do not detect the presence of WR stars, which cannot only be
attributed to the relatively low S/N of the spectrum.  Indeed, Fig.~3
compares the spectrum of the SN region with that of the WR region if it
was observed at the same S/N. The derived upper limit on the
luminosity of the He{\sc ii}4686 line (see Table~1) indicates fewer than or
 1~WR star in that region. This is not surprising given the small
number of massive stars and the relatively low WR/O star ratio, WR/O
$\sim$ 0.05, observed in the WR region.
Recall also that HST/STIS (see Fig.~1) is able to resolve the SN
region in 7 small sub-areas, and that the brightness at the precise
location of the SN is small compared to other sub-areas. 
According to the photometry of \citet{fynbo00}, in April 2000, 
the SN contributed to only 13\% of the V luminosity of the region.
Our spectroscopic observations were made 3 years after the
STIS observations and the SN remnant probably has faded away since April 2000: 
it is probable that no massive stars are present in the SN
sub-region. 

\begin{figure*}
\centering
\includegraphics[width=10cm]{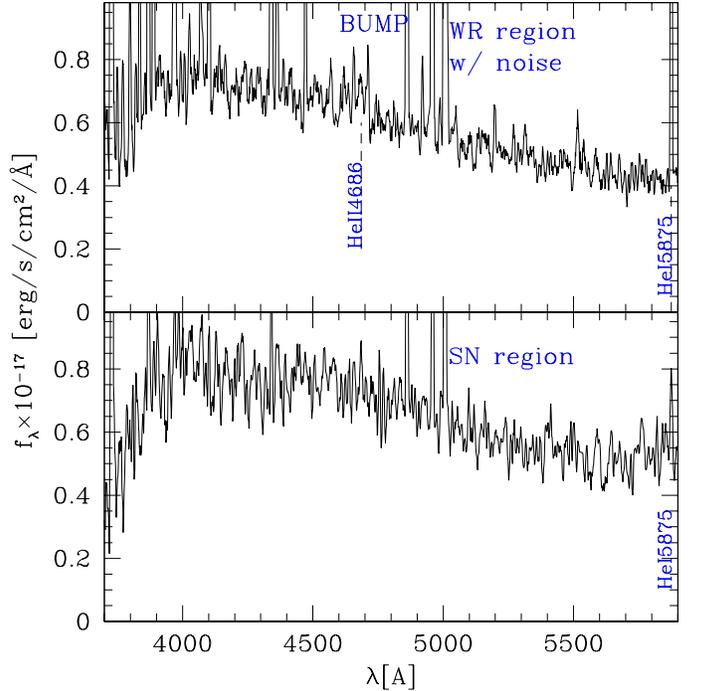}
\caption{Spectrum of the SN1998bw remnant region (bottom) compared to
	the scaled spectrum (top) of the WR region at the same S/N of
	the GRB980425 host. It shows that the absence of WR
	features in the SN1998bw remnant cannot be related simply to the
	lower S/N.  }
\end{figure*}

\subsection{GRB020903 host galaxy}

The GRB occurred in the outskirts of a small, very irregular
galaxy at $z = 0.25$ with $M_B$(AB) $= -19.3$ (see Fig.~4). The slit was centered on the
position of the reported afterglow \citep[see][]{soderberg04a}. 
 GRB020903 presents many
similarities with GRB980425. Indeed, \citet{soderberg05} have
convincingly shown that the event was followed by a supernova of
the same type as SN1998bw.  The residual HST image in Fig.~4 (top-right)
shows the precise location of the SN, which has occurred at a 0.115 arcsec
offset from a compact, unresolved region, in the outskirts
of the host galaxy.  At the distance of the host galaxy ($z = 0.25$),
this offset corresponds to  460 $\pm$100 pc, a value close to
the distance between the WR and SN regions in the GRB980425 galaxy.

\begin{figure*}
\centering
\caption{SEE ATTACHED JPG FIGURE - {\bf Top panels:} HST/ACS/F606W image taken 91 (left) and 300
	(middle) days after the GRB020903 and showing the
	associated SN \citep[same type as SN1998bw, see][]{soderberg05}. 
	The image on the right shows the residual derived from our extraction, and confirms the result of
	Soderberg et~al. The SN (peak identified with the light cross)
	exploded 460 pc off a luminous region, which is not
	resolved even at the resolution of the ACS. {\bf Middle
	panels:} HST/ACS/F606W image taken 8 (left) and 25 (middle)
	months after the GRB030329  event and showing the associated
	SN (right, light cross).  {\bf Bottom panels:} GRB980425 image
	at $z=0.0086$ (left) plotted as it would be seen if it was at
	the redshift of GRB030329 (middle) and of GRB020903
	(right). This illustrates that the offset between the SN and
	WR regions in GRB980425 is similar to that observed in
	the two more distant GRBs.  
	}
\end{figure*}

The spectrum at the position of GRB020903 (see Fig.~2) reveals an
active star forming region, with
a strong oxygen deficiency of $Z= 0.19$~$Z_{\odot}$. H$\alpha$ and
H$\beta$ show large (rest-frame) equivalent widths, 245 and 40 \AA,
respectively. These values are intermediate between
those of the SN and WR regions of the GRB980425 host. This is not
surprising, knowing that the slit has likely included several regions
around the GRB location. After correcting for extinction, we find a
luminosity of $6.9\times 10^{39}$ erg~s$^{-1}$ in the H$\beta$ line,
which can be interpreted as produced by $\sim$ 1300 O stars (assuming
a O7V stellar type) present in the aperture. A larger value might be
inferred
if the aperture correction (factor 6) was applied.

Figure~2 shows a strong signature of a blue WR bump and using the 
He{\sc ii}4686 emission line, we infer a WNL/O ratio ranging from 0.14
to 0.2 (approximately $\la$ 200 WR stars). The blue bump does
not show the same features as that of the GRB980425 WR region.
This is not so surprising after examining the large variety of blue
bump features from a survey of WR galaxies \citep{guseva00}. 
The WR bumps are made of blends of a large variety of emission
lines and can thus be subject to variations from one galaxy to
another. The relatively low S/N of our spectrum can also alter 
the appearance of the blue bump.  It is more difficult to
assess the age of the stars near the location of the GRB, because the
slit likely includes components from other regions of the
galaxy. Assuming the extinction (A$_V = 0.76$) derived from the
ionised gas, the fit of the spectral energy distribution reveals the
need for an older stellar population.

In summary, the properties of the GRB020903 host galaxy seem to be
very similar to those of the GRB980425 host, except that it shows a
higher electron
temperature and a smaller oxygen abundance. As for GRB980425, the
supernova associated with GRB020903
seems to have occurred at several hundred parsecs from a bright,
relatively 
compact region, which is most likely responsible for the numerous WR
and O stars seen in our spectrum.

\begin{figure*}
\centering
\includegraphics[width=10cm]{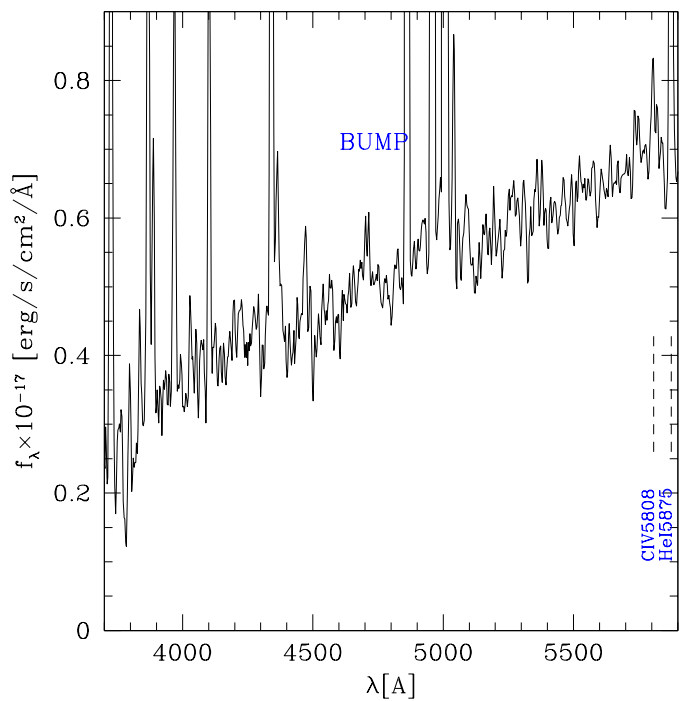}
\caption{Spectrum of the GRB031203 host ($2\times 1800$~s) taken in
	September 2004 using the 300V grism with VLT/FORS2 (from ESO
	archive, programme No. 073.D-0255(A)). The red slope of the
	energy distribution can be attributed to the low Galactic
	declination ($b^{II} = 4.7$) leading to an extinction A$_V =
	3.62$ \citep[see][]{prochaska04}. A blue bump around the
	He{\sc ii}4686 line is detected as well as a bump around the C{\sc iv}5808
	line, confirming the presence of WR stars in the host galaxy.}
\end{figure*}


\section{Discussion}

\label{s_run}

The proximity of the two GRB host galaxies (980425 and 020903) has
allowed us to detail at an unprecedented level their spatial and
spectral properties. However, these two GRB events
are among the least energetic GRBs \citep[see][]{soderberg04b}, and in
the following discussion one should be cautious before
extrapolating their properties to the numerous cosmological GRBs
observed at large distances. However, it has been argued that
sub-energetic bursts are simply events viewed away from the jet
axis. Indeed, \citet{guetta04} have shown that this
interpretation matches the statistics of both low
energetic GRBs at low redshifts and high energetic GRBs at
cosmological distances.
\citet{fynbo04} also argue for such a case for the X-ray flash XRF 030723.

\subsection{Inferences on the connection between GRB and star formation}

	GRBs occur in galaxies that are generally sub-L* and even dwarfs 
\citep[see][]{lefloch03}. The spectra of their hosts include a large
variety of emission lines, including most of the Balmer series and
Helium lines. This is characteristic of a very young burst, which is
further supported by the fact that few Myr old stars (including WR
stars) dominate the continuum.  Within these small galaxies, we find that
GRBs (and their associated supernovae) do not occur in
regions containing a large number of massive stars.
This is in strong contrast to simple expectations from the massive
star collapsar model, where the GRB progenitor star is thought to
represent a rare case among a large number of massive stars.

In both cases studied here, we find that the GRB occurred at distances
of $\sim$ 400 to 800 pc from a compact and luminous region, which is
just resolved (5~pc FWHM) in the case
of the WR region in the GRB980425 host.
In other words, GRBs neither occur directly in massive star
forming regions nor in massive LIRGs \citep[see][]{lefloch06}, but
in regions that show very few or no massive stars (e.g.\ the SN
region in the GRB980425 host). This, together with the generally
sub-solar metallicities measured in GRB hosts 
casts serious doubt on the direct relation between GRBs and
star formation, which is dominated by stellar formation in massive 
\citep[$1 < z < 3$, see][]{caputi05} or in intermediate massive 
\citep[$0.4 < z < 1$, see][]{hammer05} galaxies.
However, as discussed below, our finding of a close spatial
association and possibly a dynamic connection between GRBs and a
``nearby'' super star cluster or a massive star-forming region,
suggests a more indirect connection between GRB and star formation.


\subsection{Inferences on GRB progenitors}

What mechanisms can explain these powerful events~?
Recall that the collapsar model favours a WR origin for the GRB 
\citep{woosley99,meszaros02}. This is supported by the
specific nature of the associated supernovae, whose type (SNIb or c)
indicates that they have lost their hydrogen and/or helium envelope,
as is expected if they originate from
WR stars \citep[e.g.][]{hirschi05}. This is consistent
with the strong Nitrogen excess we observe in the SN1988bw remnant
region, which may be produced by the progenitor wind or the SN of an
initially fast rotating star \citep[see][]{meynet02}.

Are WR stars found in all GRB host galaxies~? We have investigated archival data from ESO telescopes, to acquire good spectra of the most nearby GRB host
galaxies, i.e.\ those for which 8 meter class telescopes are able to
detect the faint signature of this stellar population. Considering the
four similar objects pointed out by \citet[][ see their Fig.~2]{soderberg04b}, 
this includes our two targets plus GRB031203 ($z= 0.1055$)
and GRB030329 ($z=0.169$). Altogether these objects represent the low
energetic end of GRBs, and for all of them, supernovae have been
observed few weeks after the GRB. Nevertheless, GRB020903, GRB030329,
and GRB031203 are almost 100 times brighter than GRB980425 and are
intermediate between this event and cosmological GRBs.  Unfortunately available VLT
data on the GRB030329 host are of insufficient quality
\footnote{We extracted deep observations made at VLT in excellent seeing conditions (0.5 arcsec); unfortunately this very compact object has been observed with a wide slit (1.3 arcsec) and the  region near the bump is highly contaminated by residuals of the strong O{\sc i}5577 sky line.}.
On the other hand, the spectrum of the GRB031203 host has a
quality similar to our spectrum of GRB020903. Figure~5 shows its
spectrum and we can see that a blue bump around the He{\sc ii}4686 line and
also a bump around the C{\sc iv}5808 line are unambiguously detected,
indicating the presence of WR features, respectively of the WNL and WC
stars. For all the observed GRB hosts having a spectrum with enough quality to detect WR stars, we indeed detect WR stars.  

Our observations of the WR region of the GRB980425 host show a very
compact (5~pc FWHM) super star cluster containing thousands of O stars
and almost one hundred WR stars.
The bright region in GRB020903 separated by a projected distance of
$\sim$ 460 pc from the SN (see Fig.~4) is probably responsible for at
least a part of the observed WR emission.  It also includes a compact
component, apparently not resolved by the ACS camera and therefore
smaller than 100 pc at $z=0.25$. Unfortunately there is no HST
imagery of the GRB031203 host, and the precise location of the GRB/SN
is still unclear from ground-based data (GRB from 0.2 to 1 arcsec off
the host galaxy center according to Gam-Yan et~al. 2004 or to
Prochaska et~al. 2004). Images of the GRB030329 host 
\citep[see][]{fruchter03} reveal a very faint host galaxy ($M_V
= -16.5$) and the SN is found at its edge, off by 750 pc from the galaxy center (see Fig.~4).
The offsets revealed by Fig.~1 and 4 are very robust because they are based on HST images taken at different epochs revealing the SN, and the accuracy is better than half an ACS pixel (0.05 arcsec). They do not differ much from that found by \citet{bloom02} on a larger sample of GRBs (median value of 1.3 kpc for the offset), although their result is less accurate because it is based on a comparison between HST and ground based images.  Thus it appears that GRBs often (or always~?) occur in the outskirts of regions populated by hot and massive stars including WR
stars.

 Based on this new information on the spatial location and the stellar populations
we now discuss the possible scenarios for GRB progenitors. Assuming 
that GRB980425 is a prototype of other GRBs discussed in this paper, we are 
left with two hypotheses: either the GRB progenitor is born in a low stellar density region (in situ hypothesis) or it has been expelled from high density stellar regions (runaway hypothesis).

\subsection{GRB as runaway, fast rotating massive stars expelled from superstellar clusters ?}

If GRB hosts include clusters with WR stars in large numbers, why does
the SN (and then the GRB) occur several hundreds of parsec away from
the WR region~?
Assuming this is not a coincidence we suggest the following ``runaway
GRB'' scenario, which may be essential to achieve the necessary
conditions for the formation of a GRB. Within the high stellar density
of massive super star clusters, some stars (or double stars) can be
ejected dynamically after one or more elastic collision or from
supernova kicks in binary systems. During such a collision, the star
may have acquired a very large angular momentum, enough to lose its
hydrogen or even helium envelope providing, typically $\sim$ 3 Myr
later, a progenitor of a SNIbc such as those observed after
the GRB event.

In the case of GRB980425, for the progenitor to travel 800
pc from the WR region to the SN1998bw region in say 3--6 Myr, a velocity
of $\sim$ 260--130 km~s$^{-1}$ is necessary. 
Such velocities are somewhat larger than (but not unseen) the typical velocity
of Galactic O to early B-type runaway stars, which are thought to be runaway stars
from stellar clusters \citep[see e.g.][]{blaauw93,tenjes01,wit05}. 
This leaves this possibility for sufficiently young, i.e.\ massive GRB progenitors. 


We do not suggest that {\em all} massive runaway stars give rise to a
GRB, as this would correspond to $\sim$ 10--15\% of all O stars,
based on the knowledge of Galactic runaway stars \citep{gies87,wit05}.
However, in the case of the very massive and compact super star
clusters observed in the GRB host galaxies, the dynamical conditions,
ejection probabilities, 
and the resulting properties of runaway stars
may be quite different from the less dense and less populated
associations and clusters typically found in our Galaxy
\citep[see e.g.][]{leonard90,leonard91,portegies99}.
Furthermore, only the
runaway stars with some peculiar ejection history and high angular momentum may
produce a GRB.
An interesting case possibly resembling our suggested runaway
stars is the high velocity star HIP 60350 thought to be dynamically
ejected from the compact region NGC 3603 \citep{tenjes01}. 
 However, this object is of spectral type B3-4V, i.e.\ a priori of too low mass 
 to produce a collapsar
via ``normal'' single star evolution. \citet{hobbs05} also find large velocities for the proper motions of 
233 pulsars with mean 1D speeds of 152 km~s$^{-1}$ ($\sigma$ = 265 km~s$^{-1}$).

Runaway stars can be produced either by supernova explosions
in massive close binary system or via strong dynamical
interactions in young star clusters \citep{blaauw93}. 
It is beyond the scope of this paper to differentiate between these alternatives.  
\citet{portegies99} have simulated the ``ecology''
of a similar system, the central R136 cluster of 30 Doradus in the
Large Magellanic Cloud. They found that physical collisions between
stars are quite frequent, and closely linked with the evolution of the
star cluster. Portegies Zwart et~al. (1999, see also Leonard 1995)
argue that dynamically ejected runaway stars can be massive and could
have acquired a large angular momentum which might be consistent with
expectations for a SNIbc progenitor.  
\citet{bally05} suggest that the merger of two massive stars within dense 
clusters could be a pathway leading to hypernovae/GRBs. 
Combining our empirical findings with their models suggests
that GRBs originate from (very) rapidly rotating stars -- resulting from previous
 stellar collisions in the cluster core -- which are ejected during a subsequent
dynamical interaction.
Such dynamical interactions might also help to alleviate the
difficulties to form GRBs in rotating stellar models including the
breaking by magnetic fields found in particular at solar or somewhat
subsolar metallicities \citep{petrovic05,woosley06,yoon05}.
Most likely the runaway GRB progenitor ejected from the cluster is a single star
or a tight binary.

This scenario has several advantages.
It explains (by construction) the observed spatial shift between the GRB position and 
a nearby massive star-forming region, as observed for GRB980425 and also tentatively indicated for
GRB030329 and GRB020903. It accounts for the fact that all our studied GRB host galaxies are indeed WR galaxies.
Furthermore, it allows us to reconcile single star collapsar models with the lack of 
massive stars observed in the immediate vicinity of GRB980425.
Indeed, a rather low branching ratio of GRB/SNIbc
\citep[typically $R=N({\rm GRB \,SNe})/N({\rm SN Ibc}) \sim (2-4) \times 10^{-3}$,][]{putten04,podsi04}
implies
that statistically a GRB should be accompanied by a large number of massive stars,
which are not found in this region. In other words, if GRBs result from the
tail of a distribution of properties of massive single stars, the population 
corresponding to the remainder of this distribution, i.e.\ several thousand massive stars, 
should be present.
The age and the metallicity of the super star cluster (SSC) 
is compatible with massive single star progenitor models \citep[e.g.\ ][]{hirschi05}.

On the other hand, if the progenitor of SN1998bw was ejected from the nearby 
SSC (``WR region'' in Fig.~1) in a random direction, why is it observed within 
a small, but relatively
inconspicuous region surrounded by 6 point sources \citep[see Fig.~1 and][]{fynbo00}? 
Using Fig.~1 we have tested if this spatial association is purely fortuitous, 
assuming that the GRB progenitor is a runaway star expelled from the WR region. 
 We have drawn a circular annulus centered on the WR region with an 
 internal radius of 0.55 arcsec (to exclude the whole WR region) and an external 
 radius of 5.3 arcsec (to include the SN region); it corresponds to an area of 
 87.3 arcsec$^{2}$.  Assuming that the runaway star has been expelled in a 
 random direction, we calculate the probability of the association 
 of GRB980425 with the SN region. We find that 9.6 arcsec$^{2}$ among 87.3 
 arcsec$^{2}$ correspond to pixels with larger surface brightness than the SN 
 region, i.e. to denser stellar regions. It leads to a marginal probability 
 (11\%) of finding an expelled runaway star in a region as bright or brighter 
 than the SN region. A similar but more complex estimate can be attempted by 
 accounting for all the point sources found in the corresponding area. It is, however, 
 limited by several bright regions which could not be resolved into individual sources. The number of point sources in the 87.3 arcsec$^{2}$ area ranges from 250 and 300.  The probability of finding the expelled star surrounded by 6 point sources within 1 arcsec$^{2}$ ranges from 7 to 13\%. We conclude that the location of the SN is  marginally fortuitous and thus not unexpected.

\subsection{GRBs from binaries occurring in situ within low stellar density regions ?}

If the GRB progenitor was born in a low density region, it is rather implausible that it is a single, fast rotating, massive star.
Alternatively, several binary scenarios have been proposed for the formation of long duration
GRBs \citep[e.g.][]{fryer99}.
Ages of $\sim$ 6--8 Myr (see section 3.1.2) are a priori compatible with binary collapsar or He star merger models \citep[see][]{fryer99}.

A (massive) binary scenario does not {\em require}
a very massive, rich cluster or region, as the low GRB/SNIbc branching ratio 
might e.g.\ be  explained by processes related to the nucleation of black holes 
\citep{putten04}. However, there is no reason why in this case GRBs should be 
found preferentially in low 
stellar density regions (e.g.\ less than 10 O stars in the SN region of 
GRB980425). When considering larger samples, one should then find GRBs spatially 
distributed proportionally to the number of massive stars, i.e.\ to current star formation.

\subsection{Further observational tests}

Our observations and the two scenarios discussed above imply that
the ``classical'' single star collapsar model has to be abandoned for a binary scenario or
for our newly suggested ``runaway ejection scenario'' (or a combination of both).
What additional and possibly decisive tests distinguishing these scenarios can be envisaged?

The most decisive test of the runaway ejection scenario would be the 
localisation of {\em more isolated GRBs together with a ``parent'' super star cluster from which they were ejected.} The lack of a fairly massive and young cluster in reasonable proximity
of the GRB would exclude our scenario. We are also puzzled by the fact that, at least in two cases (GRB 980425 and GRB030329, see Figure 4), the GRB/SNs seem to be located in the prolongation of the elongated WR regions. Is this consistent with a runaway scenario ? Such investigations require the detection of many nearby GRBs, preferentially at z$<$ 0.05, which is challenging.
%

Other potentially interesting constraints may come from the study of absorption 
lines originating from the burst environment 
\citep[cf.][]{schaefer03,mirabal03,starling05,berger06,prochaska06}.
For instance, narrow Fe~{\sc ii}$^*$ and Si~{\sc ii}$^*$ fine structure absorption lines have recently 
been observed in GRB051111, indicative of a high density medium in close 
proximity to a SSC or to the GRB \citep{berger06,prochaska06}.
\citet{berger06} argue that the  source of radiative pumping required 
for the excitation of the fine structure levels must be located close to the 
absorbing medium on the GRB line of sight.
If this exciting source is a SSC, as they propose, the GRB is supposed to be 
located within the cluster --~in contrast with our observations~--, or there 
must be a chance alignment of the GRB and the SSC with its absorbing medium, 
which seems quite unlikely. Alternatively, \citet{prochaska06} argue that 
the absorbing medium could be circumstellar gas around the GRB progenitor star. 
Maintaining such a high density circumstellar envelope around a single star ejected with high velocity 
seems quite improbable. If such features turned out
to be ubiquitous, our suggested scenario would probably require the ejection 
of a tight binary rather than that of a single star.


Clearly more detailed observations including accurate spatial information 
are needed to progress on these issues.

\section{Conclusion}

We have reported here the detailed properties of the environment of
the four closest GRBs known to date, all of them being associated 
with SNe. For three of them (GRB980425,
GRB020903, and GRB031203) we have spectra of sufficient S/N allowing us
to examine the presence of WR stars, and all of them show the
characteristic WR bump. HST images are also available for the
GRB980425, GRB020903 and GRB030329 hosts. 

WR galaxies are thought to be related to very recent starbursts
($\le 10$ Myr). It is a very heterogeneous class of
galaxies, occurring in various environments \citep{schaerer98}. 
Following the \citet{conti91}
classification (more than 100 WR stars per WR galaxy) the GRB hosts
studied here are WR galaxies. \citet{smith91} estimated that 10\,\% of
HII galaxies are indeed WR galaxies. This somewhat supports the link
between GRB and WR stars, i.e.\ GRBs with very recent formation of
massive stars.

However, GRBs occurred at several hundreds of parsec away from the massive star
forming regions hosting large numbers of O and WR stars. We notice
that this applies to all the GRB hosts for which we have enough
spatial resolution (see Fig.~4). Assuming this is
purely a coincidence is not very plausible and would leave open the
question of the formation of GRB/SNIbc in a medium containing only a
small number of massive stars (e.g.\ less than 10 O stars within a
radius of 100 pc around the SN1998bw region). We believe that such a
coincidence is not fortuitous and 
propose therefore a new scenario, where GRBs are associated with
runaway stars ejected from regions showing strong and recent star
formation.

Our observations support the collapsar model by revealing for the
first time the link between WR stars and GRBs associated with SNIbc,
although in a less direct way than could have been expected. In 
our proposed scenario the frequency/rareness of the GRB events could
 be related to dynamical processes occurring during the cluster ejection.
It is also very likely that additional parameters, such as metallicity and others
suggested earlier, play a role in establishing the necessary conditions
for a GRB event. The location of GRBs in sub-luminous
galaxies and the absence of GRBs in massive galaxies 
\citep{lefloch03,lefloch06,fruchter99}, as well as the measurement of
sub-solar metallicities in GRB hosts and nearby regions 
\citep[e.g.][ and this paper]{prochaska04}, show that
metallicity also plays a role. This question will be addressed
in more detail in a subsequent paper.

%


The discovery of a possible link between runaway WR massive stars and
GRBs requires further confirmation. Massive binary scenarios must also be explored 
and tested. It has to be understood if such mechanisms occur preferentially in regions with a low density of stars or not.
 We hope that further nearby GRBs
will be soon discovered by Swift and other satellites. Follow-up
should include very high resolution images from the HST at
several different epochs, and deep spectroscopy of the region hosting
the GRB. X-shooter at VLT, equipped with an integral field unit will
be ideal for the latter purpose.

Our results illustrate the need for a better understanding of the 
physics of massive stars and the importance of dynamical processes 
in cluster environments.
It is still unclear whether the above results
support the use of GRBs to test cosmological models.


\begin{acknowledgements}

The authors have benefited from interesting discussions with various 
colleagues including
Francois Spite, Georges Meynet, Frederic Royer, Cristina Chiappini, Marc Freitag, and 
Willem-Jan de Wit. 
Gurvan Bazin contributed to a preliminary preparation of this work.
We thank the referee for thoughtful and constructive comments and questions.
Part of this work was supported by the Swiss National 
Science Foundation.
  
\end{acknowledgements}

\end{document}